\documentstyle[aps,pre,multicol,epsf]{revtex}

\begin{document}

\draft

\title{A Kac-potential treatment of nonintegrable interactions}

\author{Benjamin P. Vollmayr-Lee$^{1,2}$ and Erik Luijten$^3$}

\address{$^{1}$Department of Physics, Bucknell University, Lewisburg, PA 17837,
         USA\thanks{Permanent address.}}
\address{$^{2}$Institut f\"ur Physik, WA 331, Johannes Gutenberg-Universit\"at,
         D-55099 Mainz, Germany}
\address{$^{3}$Institute for Physical Science and Technology, University
         of Maryland, College Park, MD 20742-2431, USA\thanks{The work
         described here has been initiated at the Institut f\"ur Physik,
         Johannes Gutenberg-Universit\"at, Mainz, Germany.}}
\date{\today}

\maketitle

\begin{abstract}
  We consider $d$-dimensional systems with nonintegrable, algebraically
  decaying pairwise interactions. It is shown that, upon introduction of
  periodic boundary conditions and a long-distance cutoff in the interaction
  range, the bulk thermodynamics can be obtained rigorously by means of a
  Kac-potential treatment, leading to an exact, mean-field-like
  theory.  This explains various numerical results recently obtained for finite
  systems in the context of ``nonextensive thermodynamics,'' and in passing
  exposes a strong regulator dependence not discussed in these studies.  Our
  findings imply that, contrary to some claims, Boltzmann--Gibbs statistics are
  sufficient for a standard description of this class of nonintegrable
  interactions.
\end{abstract}


\begin{multicols}{2}

\section{Introduction}
\label{sec:intro}

In studies of critical phenomena the range of the pairwise interaction that
couples the degrees of freedom is an important consideration.  For interactions
which decay algebraically at large distances, three classes of critical
behavior may be obtained. With the standard notation $u(r) \sim
-1/r^{d+\sigma}$, one finds for system dimensionality $d<4$ that the
criticality may be characterized as {\em short-range\/} for $\sigma >
2-\eta_{\rm sr}$, non-classically {\em long-range\/} for $d/2 < \sigma <
2-\eta_{\rm sr}$, and classically {\em long-range\/} for $0 < \sigma < d/2$,
where $\eta_{\rm sr}$ is the correlation-function exponent in the corresponding
system with short-range interactions~\cite{fisher72b,sak73}.  The critical
behavior matches at the bordering cases (e.g., $\sigma \to d/2$ from above and
below) with additional logarithms.  However, for $\sigma \leq 0$ the
interactions are {\em nonintegrable}, i.e.\ $\int d^dr\, u(r) \to \infty$, and
so, under standard definitions, the thermodynamic limit does not exist.  (See
Refs.~\cite{ruelle63,fisher66,fisher70} for rigorous treatments.)

Nevertheless, recent studies have focused on this nonintegrable regime,
typically using a finite system size to render the total system energy finite.
These results are then interpreted as ``nonextensive
thermodynamics''~\cite{tsallis}, in which the system energy {\em density\/}
scales with some positive power of the system size, as do intensive variables
such as temperature. Examples of such works include molecular-dynamics
simulations of two- and three-dimensional systems with variants of the
Lennard-Jones potential~\cite{grigera96,curilef99}, Monte Carlo simulations of
one- and two-dimensional Ising~\cite{cannas96,sampaio97} and
Potts~\cite{cannas99} systems, and a numerical study of the $XY$
chain~\cite{tamarit99}.  Monte Carlo simulations have also suggested classical
critical behavior for a stochastic cellular automaton with long-range
interactions in the regime $\sigma \leq 0$~\cite{cannas98}.  On the basis of
these numerical studies of finite systems, several authors have {\em
conjectured\/} that ``nonextensive criticality'' should be
classical~\cite{cannas96,cannas99,tamarit99}.

Here we present an alternate approach to nonintegrable interactions, quite
similar to, and in certain cases equivalent to, the use of the Kac potential.
We introduce a long-distance cutoff in the interactions at some finite
distance~$R$, which enables us to use periodic boundary conditions and thus
consider homogeneous systems.  We then find the energy density in the
thermodynamic limit to scale as a power of $R$ rather than of the system
size~$L$, that is, we maintain extensitivity.  By multiplying the pair
interaction by the appropriate negative power of $R$ we recover a well-defined
$R \to \infty$ limit.  In this way, by using standard methods (including the
conventional canonical ensemble), we find an exact solution for the free energy
for all $-d\leq \sigma \leq 0$, and so demonstrate the classical nature of the
criticality without resort to simulations or conjecture!  These results have
been announced previously in Ref.~\cite{comment}.

Indeed, this is not surprising---the ``infinitely long-range and infinitely
weak interactions'' route to an explicit, analytic (mean-field-like) theory is
well known~\cite{kac59,baker61,kac63,lebowitz66}.  What is new here is the
connection between nonintegrable interactions and the much-studied
``nonextensive thermodynamics.''  This connection rests on an additional
result, namely the demonstration that {\em any\/} ordering of the limits $L,
R\to \infty$ yields the {\em same\/} free energy.  For the limit $L \to \infty$
first, the free energy is obtained directly from a Kac-potential treatment (at
least for $\sigma < 0$; the case $\sigma=0$ is treated separately), while when
the limit $R\to\infty$ is taken first, we obtain, as an intermediate step, a
finite system with constant interactions proportional to $1/L^d$---i.e.,
without approximation we obtain mean-field-like interactions!  Finally, in the
limit $R \propto L \to \infty$ the $R$-dependent prefactor multiplying the pair
interactions may be regarded instead as a power of $L$.  This reproduces all
scaling results of nonextensive thermodynamics of which we are aware, reducing
the study of these systems to the application of standard techniques with
Boltzmann--Gibbs statistics.  Furthermore, in the context of thermalized
gravitational systems~\cite{thirring80,messer81,pflug80} our
results provide what we believe is a new, direct connection to a rigorous
Kac-potential treatment.

While a finite free-energy density is obtained in the $R, L \to \infty$ limit,
the actual result depends explicitly on the cutoff {\it function}.  To be
specific, consider a fluid of density $\rho$ with the pair interaction
\begin{equation}\label{eq:pairint}
        u(r) = \cases{\infty & $r<a$ \cr
                \noalign{\medskip}
                \displaystyle
                -\frac{1}{R^{d-\tau}} {w(r/R)\over r^\tau}
                & $r>a$, $0 \leq \tau < d$ \cr
                \noalign{\medskip}
                \displaystyle -{1\over\ln R} {w(r/R)\over r^d}
                & $r>a$,
                $\tau = d$}
\end{equation}
where $\tau \equiv d+\sigma$ is introduced to avoid confusion with the negative
values of $\sigma$, and where the cutoff function $w(x)$---taken to be
isotropic for simplicity---decays at least as fast as
$1/x^{d-\tau+\varepsilon}$ for positive $\varepsilon$, with $w(0)$ finite.  The
$R$-dependent prefactors are chosen to preempt the divergence of the energy in
the limit $R \to \infty$.  From this we obtain, in the limits $L,R \to\infty$,
the Helmholtz free-energy density
\begin{equation}\label{eq:f}
        f(\rho,T) = \hbox{C.E.} \{ f_0(\rho,T) - A \rho^2 \} \;,
\end{equation}
where $\hbox{C.E.}\{\dots\}$ represents the maximal convex envelope,
$f_0(\rho,T)$ is the hard-core free energy (which is strictly proportional to
$T$), and
\begin{equation}\label{eq:A}
  A = \left\{
      \renewcommand{\arraystretch}{1.4}
      \begin{array}{lc}
        {1\over 2}S_d \int_0^\infty w(x) x^{d-\tau-1} dx & 0 \leq \tau < d \\
        {1\over 2}S_d w(0) & \tau = d
      \end{array}
      \right. \;.
\end{equation}
Here $S_d=2\pi^{d/2}/\Gamma(d/2)$ is the surface area of a unit $d$-sphere.  Up
to a factor $-k_B T$, $A$ is just the second virial coefficient without the
hard-core contribution.

The free energy thus obtained indeed depends explicitly on the cutoff function
$w(x)$, at least for $\tau < d$, but does not depend on the system shape.  We
obtain quantitatively the same result for a lattice gas (hence also for Ising
spin systems, cf.\ Ref.~\cite{hemmer76}, \S II E) with the substitution of the
lattice hard-core free energy being the only modification. In this context, we
recall that the first application of the Kac potential to spin systems is due
to Baker~\cite{baker61}.

The dependence on details of the cutoff regulator has important implications
when this solution is recast in the interpretation of ``nonextensivity.''  For
the non-periodic case of nonextensive thermodynamics, where the finite system
size is used to regulate the energy, the {\em bulk\/} quantities will depend on
both boundary effects and the system shape, a point we have not found mentioned
in previous studies.  Furthermore, when periodic boundary conditions are
employed and the interaction is cut off at some fraction of the system size,
the bulk thermodynamics will depend on precisely which fraction is used,
notwithstanding statements to the contrary~\cite{curilef99} (cf.\ our
discussion in Sec.~\ref{sec:thermodyn} below).  The remainder of the paper is
organized as follows. In Section~\ref{sec:math} we present our mathematical
treatment of the various limits which lead to the results above.  Next we
present briefly the critical properties of these systems in
Section~\ref{sec:crit}.  Since we can interpret our results in the language of
nonextensive thermodynamics, this connection is presented and discussed in
Section~\ref{sec:thermodyn}, and various difficulties with nonextensive
thermodynamics are brought to light, including the above-mentioned system-shape
dependent ``bulk'' thermodynamics.  Finally, we conclude with a summary of our
main results, and some remarks on the connection to work done in the area of
gravitational systems.

\section{Mathematical treatment}
\label{sec:math}

\subsection{Thermodynamic limit with finite range}
\label{sec:InfL}

We begin by considering a fluid in $d$ spatial dimensions with the
pair-interaction potential given by~(\ref{eq:pairint}), with finite $R$ and in
the thermodynamic limit.  Then for $0 \leq \tau < d$ our main results
(\ref{eq:f}) and~(\ref{eq:A}) for the limit $R\to\infty$ follow immediately
from the rigorous treatment of Lebowitz and Penrose \cite{lebowitz66}.  To see
this, define
\begin{equation}\label{eq:Kacphi}
        \phi(x) = w(x)/x^\tau \;,
\end{equation}
in which case~(\ref{eq:pairint}) becomes $u(r>a)= -R^{-d}\phi(r/R)$, the
canonical Kac potential, with $\phi(x)$ satisfying the necessary conditions for
the proof given in~\cite{lebowitz66}, cf.\ Eqs.\ (1.21a)--(1.21c) in this
reference.

It is also possible and useful (for the $\tau=d$ case) to understand this
result from the Mayer cluster or virial expansion about a reference hard-core
potential~\cite{hemmer76}.  Specifically, the Mayer function $\tilde
f(r)=e^{-\beta u(r)}-1$ (not to be confused with the free-energy density) may
be decomposed as
\begin{equation}\label{eq:mayer}
        \tilde f(r) = [\theta(r-a)-1] + \theta(r-a)
                [e^{-\beta u(r)} - 1] \;,
\end{equation}
with $\theta(x)$ the Heaviside step function.  The first square-bracket term is
the Mayer function for the hard-core potential and the second term accounts for
the attractions.  Each irreducible cluster of the virial expansion may be
replaced with a sum of clusters in which the individual bonds are replaced, in
turn, with each of the hard-core and attraction bonds.

The hard-core bonds are independent of the cutoff $R$, so the sum of diagrams
containing only these bonds yields the hard-core free energy (via Legendre
transformation), which is unaffected by the limit $R\to \infty$.  In what
follows, we summarize the argument for why the attraction bond 2-cluster is the
only other term which survives the $R\to\infty$ limit.

The attraction bonds go to $-\theta(r-a)\beta u(r)\propto 1/R^{d-\tau}$ for
large $R$ (with $\tau<d$, for now).  Each vertex that is free to integrate over
space gives a factor $\int d^dr w(r/R)/r^\tau \sim R^{d-\tau}$.  Consider first
diagrams with only attraction bonds: the $n$th order ring diagram (each vertex
having exactly two bonds) with $n \geq 3$ has $n$ bonds and $n-1$ free
vertices, and so vanishes as $1/R^{d-\tau}$ for large $R$.  All other $n\geq 3$
clusters have a higher ratio of bonds to vertices, and so these also vanish
when all bonds are attractive.

Replacing an attraction bond with a hard-core bond removes one factor of
$1/R^{d-\tau}$ but also kills one of the free vertex integrals, i.e.\ the
integration region is constrained to the scale of the hard core.  The net
effect for ring diagrams is that they remain of the same order in $R$, and so
vanish for $n\geq 3$.  For more complicated irreducible clusters, a situation
can arise where two vertices mutually constrained by hard-core bonds are also
connected by an attraction bond.  In this case, replacing the attraction bond
with a hard-core bond {\it does\/} increase the order of the diagram by a
factor $R^{d-\tau}$.  However, this situation can only occur for diagrams with
higher powers of $1/R$ than the ring diagrams, and can never bring them up to
order $R^0$.  Consequently, all diagrams with $n\geq 3$ and at least one
attraction bond vanish as $R\to\infty$.

The $n=2$ case is distinct as it has one bond and one free integral, so it
provides an $R^0$ contribution
\begin{eqnarray}\label{eq:two_cluster}
        a_2 &=& \lim_{R\to\infty} -{\beta\over 2} \int_{r>a} d^d r
                 {w(r/R)\over R^{d-\tau} r^\tau} \cr
                \noalign{\medskip}
        &=&  \lim_{R\to\infty} {-\beta S_d\over 2} \int_{a/R}^\infty
                         w(x) x^{d-\tau-1} dx\cr
                \noalign{\medskip}
        &=& -\beta A \;,
\end{eqnarray}
where $A$ is given in (\ref{eq:A}).

Thus the virial expansion reproduces (\ref{eq:f}), apart from the convex
envelope.  The failure of this otherwise exact method to reproduce the Maxwell
construction is that the virial expansion relies on a homogeneous density and
breaks down when this is not the case.  Nevertheless, since it is exact
whenever the density {\it is\/} homogeneous, the virial expansion supplemented
with the second law (convexity) prescribes a unique free energy, and so can be
regarded as providing the rigorous result.

The utility of the virial-expansion method is that it applies to the borderline
case $\tau=d$, where we cannot directly map to the results of
\cite{lebowitz66}.  The hard-core terms are unmodified, attraction bonds now go
as $1/\ln R$, and free vertex integrals give
\begin{equation}
\int_{r\gtrsim a} d^d r \, w(r/R)/r^d \sim \ln R
\end{equation}
The sum of diagrams with $n\geq 3$ and at least one attraction bond is $O(1/\ln
R)$ (provided there are no surviving resummations of $1/(\ln R)^n$ terms).  The
$n=2$ attraction bond diagram gives the only remaining contribution
\begin{eqnarray}\label{eq:tau_d}
        a_2 &=& \lim_{R\to\infty} {-\beta\over 2} \int_{r>a} d^d r u(r) \cr
                \noalign{\medskip}
        &=&  \lim_{R\to\infty} {-\beta S_d\over 2\ln R} \int_{a/R}^\infty
                         w(x) x^{-1} dx\cr
                \noalign{\medskip}
        &=& -{\textstyle{1\over 2}}\beta S_d w(0)
\end{eqnarray}
which again leads to (\ref{eq:f}) and (\ref{eq:A}).  Interestingly, the bulk
thermodynamics in this marginal case is not sensitive to the details of the
cutoff function.

\subsection{Infinite range with finite system size}
\label{sec:InfR}

Now consider the same fluid system, but taking $R\to\infty$ with finite $L$.
For clarity, we begin by considering a one-dimensional system with $0\leq\tau <
1$, in which the periodic boundary conditions lead to the effective pair
potential ($r>a$)
\begin{equation}
\label{eq:eff_pair_potential}
        u_{\rm eff}(r) = -{1\over R^{1-\tau}}  \sum_{n=-\infty}^\infty
                {w\bigl(|nL+r|/ R\bigr)\over|nL+r|^\tau}
\end{equation}
where $r$ is understood to be less than $L/2$.  As $R$ becomes large, the
direct ($n=0$) interaction becomes negligible and an increasingly large number
of terms contribute to the sum. By use of the expansions
\begin{equation}
\label{eq:exp-denom}
        |nL + r|^{-\tau} = |nL|^{-\tau} \left( 1 - {\tau r\over nL}
                + {\tau(\tau+1) r^2\over 2n^2L^2} - \dots \right)
\end{equation}
and
\begin{eqnarray}
        w\left(|nL + r|\over R\right) &=& w(|n|L/R) + \hbox{sign}(n){r\over R}
             w'(|n|L/R)
                \nonumber\\ && +  {r^2\over 2 R^2}w''(|n|L/R) \pm \dots ,
\end{eqnarray}
where we assume $w(x)$ to be analytic, we may rewrite $u_{\rm eff}(r)$ for
large~$R$ as
\begin{eqnarray}
\label{eq:u_asymp}
        u_{\rm eff}(r) &=& -\frac{2}{R}
               \sum_{n=1}^\infty \left(\frac{R}{nL}\right)^\tau
               \left\{ w\left(\frac{nL}{R}\right)
                \left[ 1 + {\cal O}\left(\frac{1}{n^2}\right)\right] \right.
             \nonumber \\ && \left.
             + {\cal O}\left(\frac{L}{nR},\frac{L^2}{R^2}  \right)
               \right\} \;.
\end{eqnarray}
The convergence of this series is guaranteed by the shape of the cutoff
function~$w$ and its derivatives.  Note that the leading sum for large $R$ is
{\em independent\/} of the spatial separation~$r$; the correction terms are
down by a factor $(L/R)^{1-\tau}$. In the limit of large~$R$, we may, in turn,
express this sum as an integral
\begin{eqnarray}
\label{eq:u_asymp_integral}
    \lim_{R\to\infty}  u_{\rm eff}(r) &=& -\frac{2}{R} \int_{0}^\infty
        (R/nL)^\tau  w(nL/R) \, dn
              \\
             &=& -\frac{2}{L} \int_0^\infty w(x) x^{-\tau} \, dx \;.
\end{eqnarray}

For the borderline case $\tau = 1$, Eqs. (\ref{eq:eff_pair_potential})
and~(\ref{eq:u_asymp}) have to be multiplied by $(\ln R)^{-1}$.  Furthermore
Eq.~(\ref{eq:u_asymp_integral}) now requires a lower integration limit $n=1$
(or any constant) as a regulator.  The pair interaction is then
\begin{eqnarray}
  u_{\rm eff}(r) &=&  -\frac{2}{L \ln R} \int_{L/R}^\infty w(x) x^{-1} \, dx \\
       &=&  -\frac{2}{L \ln R}[ w(0)\ln R + O(R^0) ] \;.
\end{eqnarray}
In the limit $R \to \infty$, we find the effective pair potential is again
independent of the spatial separation, $u_{\rm eff}(r) = -2L^{-1}w(0)$.

The generalization of this treatment to higher dimensionalities is
straightforward. For a $d$-dimensional system of size $L_1 \times L_2 \times
\cdots \times L_d$, with periodic boundary conditions and $0 \leq \tau < d$,
the effective pair potential ($r>a$) is
\begin{equation}
\label{eq:eff_pair_potential-d}
        u_{\rm eff}(r) = -{1\over R^{d-\tau}}
            \sum_{n_1=-\infty}^\infty \cdots \sum_{n_d=-\infty}^\infty
                {w( \frac{\tilde r(n_1,\dots,n_d)}{R} )
                        \over \tilde r(n_1,\dots,n_d)^\tau}
\end{equation}
where
\begin{equation}\label{eq:tilde-r}
        \tilde r(n_1,\dots,n_d) \equiv | {\bf r} + n_1 L_1\hat{\bf x}_1
             + \dots + n_d L_d\hat{\bf x}_d |
\end{equation}
is the separation corresponding to the $(n_1,\dots,n_d)$ periodic repeat, and
the $\hat{\bf x}_i$ are orthonormal vectors.  In the limit of infinite
interaction range~$R$, the direct interaction term (all $n_i=0$) becomes
negligible compared to the sum and we find the following generalization of
Eq.~(\ref{eq:u_asymp_integral}):
\begin{equation}
\label{eq:u_asymp_integral-d}
 u_{\rm eff}(r) = - \frac{S_d}{V} \int_{0}^\infty w(x) x^{d-1-\tau} \, dx \;,
\end{equation}
where $S_d$ has been introduced below Eq.~(\ref{eq:A}) and $V \equiv
\prod_{i=1}^{d} L_i$. We note that the strength of this constant interaction is
inversely proportional to the volume, just as one expects for a mean-field-like
system~\cite{baxter82}. For $\tau=d$ one has, in the limit $R \to \infty$,
$u_{\rm eff}(r) = -S_d V^{-1} w(0)$.  Since a constant pair interaction gives a
configuration-independent energy density $\frac{1}{2} u_{\rm eff} \rho^2 V$,
the energy and entropy contributions to the free energy decouple, and one
obtains (\ref{eq:f}) directly in the thermodynamic limit.

Instead of a ``soft'' cutoff provided by the (analytic) function~$w(x)$, one
can also introduce a ``hard'' cutoff at a distance~$R$, which is equivalent to
taking, say, $w(x)=\theta(1-x)$ in Eq.~(\ref{eq:pairint}). We start again with
$d=1$ and $\tau<d$. The effective pair potential can be written as
\begin{equation}
  u_{\rm eff}(r) = -{1\over R^{1-\tau}} \sum_{n=-R/L}^{R/L}
  \frac{1}{|nL+r|^\tau} \;,
\end{equation}
where, for simplicity, we have taken $R$ to be an integer multiple of~$L$.  For
large~$R$, the $n=0$ term becomes negligible compared to the sum and by use of
the expansion~(\ref{eq:exp-denom}) we rewrite $u_{\rm eff}(r)$ as
\begin{equation}
  u_{\rm eff}(r) =
  - \frac{2}{L}\left(\frac{R}{L}\right)^{\tau-1}
    \sum_{n=1}^{R/L} \left\{ \frac{1}{n^\tau}
    \left[1 + {\cal O}\left(\frac{1}{n^2}\right) \right] \right\} \;.
\end{equation}
The ${\cal O}(n^{-2})$ correction terms are again down by a factor
$(R/L)^{\tau-1}$ from to the leading sum [compare to (\ref{eq:u_asymp})].  As
before, we end up with a constant effective interaction $\lim_{R \to \infty}
u_{\rm eff}(r) = - 2L^{-1}/(1-\tau)$.  For $\tau=1$, the correction terms decay
only like $1/\ln R$ and $\lim_{R \to \infty} u_{\rm eff}(r) = -2/L$.  Since the
case of general dimensionality can be treated along the same lines, we only
mention the resulting values for the constant pair potential:
$\lim_{R\to\infty} u_{\rm eff}(r) = - S_d V^{-1} / (d-\tau)$ for $\tau < d$ and
$\lim_{R\to\infty} u_{\rm eff}(r) = - S_d / V$ for $\tau = d$.  These results
are all consistent with (\ref{eq:f}) with $w(x)=\theta(1-x)$.

Thus, we have shown that, in the limit of infinite interaction range, all pair
interactions are identical even in a finite system, for any $0 \leq \tau \leq
d$. For $\tau>d$ on the other hand, the sum in Eqs.\
(\ref{eq:eff_pair_potential}),~(\ref{eq:eff_pair_potential-d}) is convergent
and hence the $R$-dependent prefactor in $u(r)$ should be omitted.  The first
term within the square brackets is then no longer negligible in the limit $R
\to \infty$, which leads to an effective interaction that depends both on $r$
and on $\tau$ and consequently to nontrivial critical behavior. Interestingly,
the critical {\em exponents\/} retain their classical values until $\tau >
3d/2$, except for the correlation-length exponent~$\nu$ and the
correlation-function exponent~$\eta$~\cite{fisher72b}.  The critical
temperature and other nonuniversal quantities exhibit a nontrivial $\tau$
dependence already for $\tau > d$, cf.\ Ref.~\cite{class-lr} and references
therein.

\subsection{Lattice gas}
\label{sec:LatGas}

So far we have only discussed continuum fluid criticality.  The generalization
to a lattice gas (and, of course, to the Ising ferromagnet) results in the same
free energy (\ref{eq:f}) with the appropriate lattice hard core $f_0(\rho)$.
This may be seen most directly in the $R\to \infty$ first case, where the pair
interactions go to a constant.  This derivation applies both for particles in a
continuum or on a lattice (indeed, no specification was made), and so the
constant interactions are obtained in both cases.  The final step of
constructing the free energy from the decoupled energy and entropy reveals that
the lattice hard-core free energy is the appropriate one to use
in~(\ref{eq:f}).

For the $L\to\infty$ first limit, one must use a lattice generalization of the
Mayer expansion.  Such an expansion would give for non-interacting particles
the lattice hard-core free energy.  The lattice sums for clusters with
attraction bonds can be taken to integrals in the large $R$ limit, and so the
rest of the continuum derivation applies.

\subsection{Thermodynamic limit with a system-size dependent range}
\label{sec:SimLim}

\begin{figure}
\narrowtext
\epsfxsize=\hsize\epsfbox{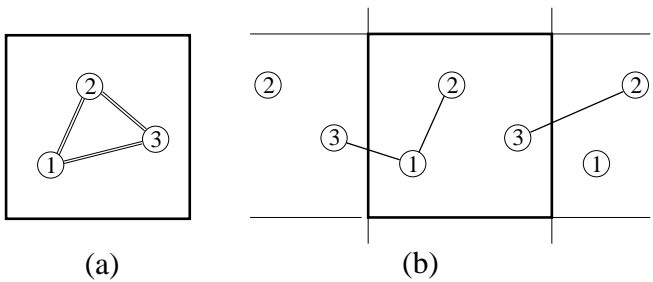}
\medskip
\caption{(a) The 3-cluster formed with the effective interaction $u_{\rm eff}$
and (b) one contribution to the decomposition of this cluster into pair
interactions $u(r)$.}
\label{fig:cluster}
\end{figure}

The final case we consider is that of $R\propto L\to\infty$, the case most
directly applicable to ``nonextensive thermodynamics.''  Consider a finite
system with periodic boundary conditions (with all dimensions $L_i\propto L$)
and direct pair interactions given by (\ref{eq:pairint}) for some finite~$R$.
The periodic boundary conditions give rise to the effective pair
potential~(\ref{eq:eff_pair_potential-d}).  The virial expansion for the
pressure may then be obtained from a finite-volume cluster expansion with the
Mayer function $\tilde f=\exp(-\beta u_{\rm eff})-1 \sim -\beta u_{\rm eff}$
for large $R$.  Since the effective interaction is a sum of pair interactions,
a simple $u_{\rm eff}$ cluster such as the irreducible 3-cluster shown in
Fig.~\ref{fig:cluster}(a) decomposes into a large number of pair-interaction
bond clusters [cf.\ Fig.~\ref{fig:cluster}(b)].  Note that, in order to avoid
overcounting, one end of every bond must remain in the original finite volume,
but the other end may be taken to lie in any of the periodic repeats (and these
still contribute to the virial expansion as they are parts of an irreducible
cluster).

In spite of these complications, we remark that all hard-core bonds are the
same as in the previous ($L\to\infty$ first) case, since they only appear in
the direct interaction of the original pair.  Hence the hard-core bonds sum to
the hard-core free energy regardless of whether $L$ and $R$ are simultaneously
taken to infinity, or $L$ first.

Furthermore, all $n$-clusters with $n \geq 3$ and at least one attraction bond
can be shown to still vanish as $R,L\to\infty$.  First, all bonds still carry a
factor of $1/R^{d-\tau}$ for large $R$, as before.  Free vertices integrate to
\begin{equation}\label{eq:vert_int}
  \int_{L^d} d^dr \, r^{-\tau} w(r/R) \sim R^{d-\tau} \int_{(L/R)^d} d^dx \,
  x^{-\tau} w(x)
\end{equation}
for the direct interaction of the original pair, which goes as $R^{d-\tau}$
times a finite factor for large~$R$.  For a pair interaction involving a
neighboring replica, the vertex integration is the same as above with a shift
in the argument of $w(\tilde r/R)$, where $\tilde r$ [see
Eq.~(\ref{eq:tilde-r})] goes roughly as $r+cL$.  The resulting vertex
integration will also scale as $R^{d-\tau}$, provided $r+cL \lesssim R$.  Since
$\tilde r$ increases for each increasingly remote replica, the cutoff function
$w(x)$ will ensure that only a number ${\cal O}((R/L)^d)$ of such bonds will
contribute.  Hence free vertex integration with effective interaction bonds,
while considerably more complicated, still results in a factor $R^{d-\tau}$ at
large $R$.  Previous arguments from Section~\ref{sec:InfL} then apply, and so
these terms all vanish when $R,L\to\infty$.

The remaining 2-cluster integral may be written as
\begin{eqnarray}
        a_2 &=&  \lim_{R\to\infty}{\beta\over 2} \int_{r>a} d^d r \,
                 u_{\rm eff}(r) \\
                 \noalign{\medskip}\label{eq:sum1}
      &\sim&  {-\beta\over 2 R^{d-\tau}}
        \prod_{i=1}^d\left(\sum_{n_i=-\infty}^\infty
        \int_0^{L_i} dx_i \right) {w(\tilde r/R)\over\tilde r^\tau} \\
      \noalign{\medskip}\label{eq:sum2}
      &=&  {-\beta\over 2 R^{d-\tau}}
        \prod_{i=1}^d\left(\sum_{n_i=-\infty}^\infty
        \int_{n_iL_i}^{(n_i+1)L_i} dx_i \right) {w(r/R)\over r^\tau} \;,
\end{eqnarray}
where the sums cover all replicas (and we have omitted writing the hard-core
condition for clarity).  In going from (\ref{eq:sum1}) to (\ref{eq:sum2}) the
integrand variable changed from $\tilde r$ to $r$, consistent with the
definition of $\tilde r$.  The remaining integrals piece together a single
volume integral over all space, so
\begin{eqnarray}
  a_2 &=& \lim_{R\to\infty}{-\beta\over 2 R^{d-\tau}}\int_{r>a} d^dr
  \, w(r/R)/r^{\tau} \cr
  \noalign{\medskip}
     &=& -\beta A \;,
\end{eqnarray}
following (\ref{eq:two_cluster}).  Combining this with the hard-core
contribution gives the same free energy (\ref{eq:f}) in the $R,L \to \infty$
limit as was found previously for $L$ or $R$ going to infinity first.  The
significance of this simultaneous limit towards nonextensive thermodynamics
will be discussed in Section~\ref{sec:thermodyn}.

\section{Critical behavior}
\label{sec:crit}

The critical properties of nonintegrable systems are readily found from the
Helmholtz free-energy density~(\ref{eq:f}) via standard procedures.  For
example, the critical density can be obtained from $\partial^3
f(\rho,T)/\partial\rho^3|_{\rho_c} = 0$, which reduces to the temperature (and
attraction) independent condition $\partial^3
f_0(\rho,T)/\partial\rho^3|_{\rho_c} = 0$.  The critical temperature is then
found from $\partial^2 f(\rho,T_c)/\partial\rho^2|_{\rho_c} = 0$ which gives
\begin{equation}\label{eq:Tc}
  k_B T_c = 2 A  \left({\partial^2 (\beta f_0)\over\partial\rho^2}
  \right)^{-1}_{\rho=\rho_c} \;.
\end{equation}

For a hypercubic lattice with lattice constant~$a$, the lattice-gas hard-core
free energy is $\beta f^{\rm LG}_0 = \rho \ln\rho +
(a^{-d}-\rho)\ln(1-a^d\rho)$.  This results in the critical values
\begin{equation}
\hbox{lattice gas:}\quad  a^d\rho_c = 1/2 \qquad k_B T_c = A/(2a^d) \;.
\end{equation}
Although the free energy for continuum hard spheres (diameter $a$) is not known
exactly, a very good approximation in three dimensions is nevertheless given by
the Carnahan--Starling (CS) expression
\begin{equation}
  \beta f^{\rm CS}_0(\rho) = \beta f^{\rm Id}(\rho) +
     {\rho^2 v_0(4-3\rho v_0)\over(1-\rho v_0)^2} \;,
\end{equation}
with $v_0=\pi a^3/6$ the volume of the hard sphere and $f^{\rm Id}$ the ideal
gas free energy density.  Using this for $f_0$ yields a quintic equation for
$\rho_c$ with a unique positive root~\cite{fisher96} and the critical values
\begin{equation}
  \hbox{CS:}\quad a^3 \rho_c\simeq 0.249129 \quad k_B T_c \simeq
  0.180155 A/a^3 \;.
\end{equation}

The susceptibility $\chi$, defined as the ratio of the isothermal
compressibility to the ideal gas compressibility,
\begin{equation}
\chi^{-1}(\rho,T) \equiv \rho {\partial^2(\beta f)\over\partial\rho^2} \;,
\end{equation}
exhibits near criticality the classical divergence $\chi \sim C_+/t$ for
$\rho=\rho_c$ and positive reduced temperature $t\equiv (T-T_c)/T_c$.  From
(\ref{eq:f}) we find
\begin{equation}
C_+ = \left(\rho_c {\partial^2(\beta f_0)\over\partial\rho^2}\bigg|_{\rho_c}
\right)^{-1} \;,
\end{equation}
which results in $C^{\rm LG}_+=1/2$ for the lattice gas and $C^{\rm CS}_+
\simeq 0.361569$ for the $d=3$ hard-sphere fluid, approximated by the CS
expression.

Finally, the order parameter for $T<T_c$ is given by
\begin{equation}
  a^d \left|\rho-\rho_c\right| = B \sqrt{-t}
\end{equation}
for $t$ sufficiently small, with
\begin{equation}
  B = a^d \left(18 [\partial^2 f_0/\partial\rho^2]_{\rho_c} \over
      [\partial^4 f_0/\partial\rho^4]_{\rho_c} \right)^{1/2} \;.
\end{equation}
This evaluates to $B^{\rm LG}=3/2$ and $B^{\rm CS}\simeq 1.13459$.

\section{Nonextensive thermodynamics}
\label{sec:thermodyn}

As indicated in the Introduction, the main motivation for this work stems from
the considerable attention systems with nonintegrable interactions have
received in the context of ``nonextensive thermodynamics.''  An essential
aspect of these studies of nonextensivity is the use of the system size as the
regulator for the energy.  Furthermore, the interactions are not scaled by a
negative power of the system size but left with strength of order unity.

Our cutoff interaction with range $R$ can be interpreted directly in terms of
this nonextensive thermodynamics for $R/L$ constant, which reproduces the
system-size regulator. The negative power of $L$ multiplying our pair
interaction is eliminated by scaling the temperature according to $T \to
L^{d-\tau} T$ in the Boltzmann factor ($T\to T\ln L$ for $\tau=d$).  The
scaling of this system-size dependent ``temperature'' (and consequently the
free energy) matches exactly the conjectures of nonextensive thermodynamics,
thus we have derived rigorously the primary conclusions of
Refs.~\cite{cannas96,curilef99,cannas99,tamarit99} using only standard methods.
We note that, despite what has been suggested in Ref.~\cite{salazar00}, neither
the explicit free energy nor exact results for the nature of the criticality
were obtained in Refs.~\cite{cannas96,cannas99}.

Next we want to discuss some examples from the recent literature and point out
a few problems attached to the interpretation of nonextensive thermodynamics.
First, a pervasive notational problem in the nonextensive thermodynamics
literature is the use of ``long-range interactions'' to mean ``nonintegrable
interactions.'' The former term already has a standard meaning within the
considerably more important class of integrable interactions
\cite{ruelle68,dyson69,fisher72b}.  The converse problem also exists, where all
integrable interactions (including the true long-range interactions) are termed
``short-range'' (see, e.g., Ref.~\cite{jund95}).

Many papers addressing nonintegrable interactions have relied on numerical
simulations of finite systems. In general, the required regularization of the
interactions is carried out either by imposing free boundary conditions (which
leads to an inhomogeneous system) or by adopting periodic boundary conditions
and cutting off the interaction at half the system size, being the maximum
separation between the particles. For example, Curilef and
Tsallis~\cite{curilef99} have performed molecular-dynamics simulations of
fluids in $d=2,3$ dimensions with Lennard-Jones-like interactions, with an
attractive tail decaying like $r^{-\tau}$ and $1 \leq \tau \leq 2d$. It is
explicitly stated (Ref.~\cite{curilef99}, p.~271) that, in the thermodynamic
limit, no physical consequences emerge from the (``computationally
convenient'') adoption of a cutoff at half the size of the simulation box.  Our
exact solution presented in Sec.~\ref{sec:math} shows that this is not correct
for $\tau<d$, but that rather different {\em bulk\/} thermodynamics emerges for
different cutoff distances and cutoff functions.

In Ref.~\cite{jund95}, rings of magnetic particles in a colloidal suspension
have been studied numerically, where the nonmagnetic part of the interactions
has the above-mentioned generalized Lennard-Jones form.  The main results are
heuristically interpreted in terms of ``nonextensive thermodynamics'' by
observing that the size dependence of the total energy of the rings can be well
described by a scaling law obtained from the integrated interaction (which is
essentially a mean-field-like approximation). This scaling law is just what is
also found within the ``$q$-generalized thermodynamics'' of
Ref.~\cite{tsallis}, commonly referred to as Tsallis $q$-statistics, and
consequently a $\tau$ dependence ($\tau=d+\sigma$) is proposed for the
so-called ``nonextensivity parameter''~$q$ appearing in this formalism: $q=1$
(corresponding to Boltzmann--Gibbs statistics) for $\tau > d$ and $q=2-d/\tau$
for $\tau \leq d$.  Also in Ref.~\cite{nobre95} it has been conjectured that
$q$ is a $d$- and $\tau$-dependent quantity for $\tau \leq d$.  It appears that
these conclusions have been since revised~\cite{sampaio97,tsallis99}, and these
systems are now classified as ``weakly violating'' Boltzmann--Gibbs statistics,
meaning that $q=1$, whereas several thermodynamic quantities lose their
extensivity.  Nevertheless, recent studies have continued to explore $q \neq 1$
values for $\tau \leq d$, citing an alleged natural connection between Tsallis
statistics and nonintegrable interactions \cite{salazar99}.  Our results show
that the same energy scaling for $\tau \leq d$ may be obtained with the
conventional value $q=1$.

Inspired by the system-size dependence of the energy found in
Ref.~\cite{jund95}, an Ising model with interactions decaying as $r^{-\tau}$
($\tau \geq 0$) and free boundary conditions has been analyzed by mean-field
methods in Ref.~\cite{cannas96}. On the basis of the resulting values of the
critical temperature for $\tau=0$ and $\tau=d$, it was then {\em conjectured\/}
that the mean-field prediction for the critical temperature might hold for all
$0 \leq \tau \leq d$. Our exact result now demonstrates that this is indeed
true, but reveals in addition an awkward consequence of the boundary conditions
adopted in Ref.~\cite{cannas96}: since the thermodynamic limit explicitly
depends on the choice of the cutoff function, an inhomogeneous system with an
inhomogeneous cutoff will lead to bulk thermodynamics that depends on the {\em
shape\/} of the system, which to our judgment constitutes an undesirable
feature of the nonextensivity formulation.  We note that already in an early
study of the Ising model with long-range interactions~\cite{hiley65} a
system-shape dependence of the thermodynamic properties has been observed for
the case of {\em conditionally convergent\/} lattice sums with~$\tau=d$ (such
as dipolar forces). Also the exactness of mean-field theory for {\em
shape-independent\/} forces in the limit $\tau \to d+$ has been obtained in the
same reference, essentially from an observation similar to ours for general $0
\leq \tau \leq d$, namely that one divergent term dominates all other terms in
the lattice sum. However, for the case $\tau < d$, we have been unable to find
in the literature any mention of the system-shape dependent thermodynamics that
must result for all inhomogeneous systems with nonintegrable interactions and a
system-size dependent cutoff.

\section{Summary and conclusions}

In summary, we have shown that $d$-dimensional, periodic systems with
nonintegrable, algebraically decaying interactions, i.e., interactions of the
form $u(r) \sim -1/r^\tau$, with $0 \leq \tau \leq d$, are exactly described by
mean-field theory, upon introduction of a cutoff~$R$ in the interaction range
and the proper $R$-dependent rescaling of the interaction strength. This proof
holds for either order of limits $R \to \infty$ and $L \to \infty$ (where $L$
is the linear system size), including the simultaneous limit, and the resulting
free energy depends on the details of the cutoff.

Our study employs Boltzmann--Gibbs statistics and pertains directly to
nonextensive thermodynamics, providing explicit, exact results for the
thermodynamics and critical behavior.  In doing so, we show that nonintegrable
interactions do not require~\cite{nobre95,salazar99} the application of
generalized $q$-statistics.  Furthermore, the explicit regulator
dependence---cutoff length, cutoff shape, and even system shape for
inhomogeneous systems---is demonstrated, a topic which has been mostly
neglected in nonextensivity studies.

On an intuitive level, our findings for the case of finite systems with
infinite~$R$ (Section~\ref{sec:InfR}) result from the divergence of the lattice
sums over the periodic copies of the system under consideration; these
divergent sums then dominate the direct pair interaction, and are, to leading
order, independent of the spatial separation between particles. A suitable
normalization is indispensable for the existence of the thermodynamic limit
and---as we have pointed out---a regulator depending on the number of
interactions rather than on the system size emerges as the natural choice.  The
resulting effective pair interaction is then independent of the spatial
separation, for $0 \leq \tau \leq d$.

We have concentrated on both fluids and lattice gases.  However, since our
method for the $R\to\infty$ first case demonstrated how the effective
interaction becomes independent of spatial separation, our results immediately
carry over to large classes of other systems as well (obviously, the critical
properties obtained in Sec.~\ref{sec:crit} explicitly refer to systems with a
one-component order parameter). These include general ${\rm O}(n)$ models
($XY$, Heisenberg, $\ldots$) and Potts models. In this context we note that in
earlier work the exactness of mean-field theory in the limit $\tau \to d+$ has
been found not only for the Ising model ($n=1$)~\cite{hiley65} but also for the
Husimi--Temperley mean spherical model ($n \to \infty$)~\cite{brankov90}.

Our explicit result of an analytic free energy, generalized to the systems
mentioned above, explains a number of numerical results obtained for systems
with nonintegrable interactions. This includes the molecular-dynamics
simulations of systems with a generalized Lennard-Jones potential discussed in
Sec.~\ref{sec:thermodyn} \cite{grigera96,curilef99}, Monte Carlo simulations of
one-dimensional Ising~\cite{cannas96}, Potts~\cite{cannas99}, and
$XY$~\cite{tamarit99} models with $\tau < 1$, and the scaling properties found
in Refs.~\cite{jund95,sampaio97,anteneodo98,cannas98}. In
Ref.~\cite{bergersen95}, a subleading term in the spin--spin correlation
function of the nonintegrable Ising chain was considered and on the basis of
Monte Carlo simulations of finite systems with $\tau = 0.50$ and $\tau = 0.75$
it was concluded that these correlations are correctly described by mean-field
theory.

Finally, we remark on the connection of these results to gravitational systems,
where $d=3$ and $\tau=1$.  Since the masses of the particles and the
gravitational coupling are presumably fixed, we are not at liberty to scale to
infinitely weak interactions.  However, following the ``nonextensive
thermodynamics'' formulations, we can consider $R\propto L$ for a finite system
and regard the prefactor in the pair interactions as belonging to a rescaled
temperature $T\to L^{d-\tau}T = L^2T$. At the same time, we note that such a
divergent temperature appears of limited practical use.

The existence of a phase transition in these systems has been studied for
fermionic particles in the context of Thomas--Fermi
theory~\cite{thirring80,messer81}.  Most studies have considered
gravitationally interacting particles in a non-periodic, finite-sized system,
concentrating on the asymptotic, large-$N$ limit.  Here they obtain
non-extensivity in the particle number~$N$, with the energy per particle
growing as $N^{4/3}$ (in contrast to $N^0$ for integrable interactions).  In
Ref.~\cite{pflug80} the energy densities were suitably rescaled to enable the
infinite-volume limit, which presumably comes closest to what has been
presented in the current work.

However, these studies differ from ours due to the fermionic character of the
particles, which is used to regulate the short-distance behavior.  Classical
gravitational systems have previously been studied as
well~\cite{messer82,kiessling89}, again in a non-periodic, finite-sized box,
with various forms of short-distance regulators.  In this case the energy per
particle scales as $N$, essentially because the pair interaction, while
decaying as $1/r$, has a minimum value for all particles that is proportional
to $1/L$, due to the fixed system size.  In contrast, our thermodynamic limit,
$L\to\infty$ with fixed particle density, combined with the long-distance
power-law tails of the pair potential, implies an energy per particle growing
as $N^{2/3}$ (or $N^{1-\tau/d}$ for general $\tau<d$).  Finally, we mention
Ref.~\cite{lynden-bell99} for a recent review of some interesting features of
thermalized gravitational systems.

\acknowledgments

We wish to thank Professor Kurt Binder for a stimulating remark, and Professor
Michael E. Fisher and Professor Bob Dorfman for helpful comments on the
manuscript.  We are indebted to Professor Michael Kiessling for enlightening us
on the connection to gravitational systems.  The hospitality of the Condensed
Matter Theory Group at the Johannes Gutenberg-Universit\"at Mainz, where this
work was initiated, is gratefully acknowledged.  B.P.V.-L. acknowledges support
from {\em Sonderforschungsbereich\/} 262. E.L. acknowledges financial support
through a fellowship of the Max-Planck-Institute for Polymer Research, from the
National Science Foundation (through Grant No. CHE 99-81772 to Prof.\ M.E.
Fisher), and from the Department of Energy, Office of Basic Energy Sciences
(through Grant No. DE-FG02-98ER14858 to Prof.\ A.Z. Panagiotopoulos).

\end{multicols}


\begin{references}
  
\bibitem{fisher72b} M.~E. Fisher, S.-k. Ma, and B.~G. Nickel, Phys. Rev. Lett.
  {\bf 29}, 917 (1972).

\bibitem{sak73} J. Sak, Phys. Rev. B {\bf 8}, 281 (1973).

\bibitem{ruelle63} D. Ruelle, Helv. Phys. Acta {\bf 36}, 183 (1963).

\bibitem{fisher66} M. E. Fisher and D. Ruelle, J. Math. Phys. {\bf 7}, 260
  (1966).

\bibitem{fisher70} M. E. Fisher and J. L. Lebowitz, Comm. Math. Phys. {\bf 19},
  251 (1970).

\bibitem{tsallis} C. Tsallis, J. Stat. Phys. {\bf 52}, 479 (1988).

\bibitem{grigera96} J. R. Grigera, Phys. Lett. A {\bf 217}, 47 (1996).

\bibitem{curilef99} S. Curilef and C. Tsallis, Phys. Lett. A {\bf 264},
  270 (1999).

\bibitem{cannas96} S. A. Cannas and F. A. Tamarit, Phys.\ Rev.\ B {\bf 54},
  12661 (1996).

\bibitem{sampaio97} L. C. Sampaio, M. P. de Albuquerque, and F. S. de Menezes,
  Phys. Rev. B {\bf 55}, 5611 (1997).

\bibitem{cannas99} S. A. Cannas, A. C. N. de Magalh\~aes, and F. A. Tamarit,
  Phys. Rev. B {\bf 61}, 11521 (2000).

\bibitem{tamarit99} F. Tamarit and C. Anteneodo, Phys. Rev. Lett.
  {\bf 84}, 208 (2000).

\bibitem{cannas98} S. A. Cannas, Physica A {\bf 358}, 32 (1998).

\bibitem{comment} B. P. Vollmayr-Lee and Erik Luijten, Phys. Rev. Lett.
  {\bf 85}, 470 (2000).

\bibitem{kac59} M. Kac, Phys. Fluids {\bf 2}, 8 (1959).

\bibitem{baker61} G. A. Baker, Phys. Rev. {\bf 122}, 1477 (1961).

\bibitem{kac63} M. Kac, G. E. Uhlenbeck and P.C. Hemmer, J. Math. Phys. {\bf
  4}, 216 (1963).

\bibitem{lebowitz66} J. L. Lebowitz and O. Penrose, J. Math. Phys. {\bf 7},
  98 (1966).

\bibitem{thirring80} W. Thirring, {\it Lehrbuch der Mathematischen Physik},
  Vol.~4 (Springer, Vienna, 1980). 

\bibitem{messer81} J. Messer, {\it Temperature Dependent Thomas-Fermi Theory\/}
  (Springer, Berlin, 1981).

\bibitem{pflug80} A. Pflug, Comm. Math. Phys. {\bf 78}, 83 (1980).

\bibitem{hemmer76} P. C. Hemmer and J. L. Lebowitz, in {\it Phase Transitions
  and Critical Phenomena}, Vol.~5B, edited by C. Domb and M. S. Green
  (Academic, London, 1976).

\bibitem{baxter82} R. J. Baxter, {\it Exactly Solved Models in Statistical
  Mechanics\/} (Academic, London, 1982).

\bibitem{class-lr} E. Luijten and H. W. J. Bl\"ote, Phys. Rev. B {\bf 56}, 8945
  (1997).

\bibitem{fisher96} M. E. Fisher and B. P. Lee, Phys. Rev. Lett. {\bf 77}, 3561
  (1996).

\bibitem{salazar00} R. Salazar and R. Toral, Phys. Rev. Lett. {\bf 85}, 471
  (2000).

\bibitem{ruelle68} D. Ruelle, Comm. Math. Phys. {\bf 9}, 267 (1968).

\bibitem{dyson69} F. J. Dyson, Comm. Math. Phys. {\bf 12}, 91 (1969).

\bibitem{jund95} P. Jund, S. G. Kim, and C. Tsallis, Phys. Rev. B {\bf 52}, 50
  (1995).

\bibitem{nobre95} F. D. Nobre and C. Tsallis, Physica A {\bf 213}, 337 (1995).

\bibitem{tsallis99} C. Tsallis, in {\it Nonextensive Statistical Mechanics and
  Thermodynamics}, edited by S. R. A. Salinas and C. Tsallis, Braz. J. Phys.
  {\bf 29}, 1 (1999).

\bibitem{salazar99} R. Salazar and R. Toral, Phys. Rev. Lett. {\bf 83}, 4233
  (1999).  Note also Ref.~\protect\cite{salazar00}, which clarifies the
  boundary conditions employed.

\bibitem{hiley65} B. J. Hiley and G. S. Joyce, Proc. Phys. Soc. {\bf 85}, 493
  (1965).

\bibitem{brankov90} J. G. Brankov, Physica A {\bf 168}, 1035 (1990).

\bibitem{anteneodo98} C. Anteneodo and C. Tsallis, Phys. Rev. Lett. {\bf 80},
  5313 (1998).

\bibitem{bergersen95} B. Bergersen, Z. R\'acz, and H.-J. Xu, Phys.\ Rev.\ E
  {\bf 52}, 6031 (1995).

\bibitem{messer82} J. Messer and H. Spohn, J. Stat. Phys. {\bf 29}, 561 (1982).
  
\bibitem{kiessling89} M. K.-H. Kiessling, J. Stat. Phys. {\bf 55}, 203
  (1989).

\bibitem{lynden-bell99} D. Lynden-Bell, Physica A {\bf 263}, 293 (1999).

\end{references}
\end{document}